\documentclass{llncs}
\usepackage{amssymb}
\usepackage{graphicx}
\usepackage{color}
\usepackage[hyphens]{url}
\usepackage{multirow}
\usepackage{rotating}
\usepackage{tikz}

\newenvironment{uriquote}{\list{}{\leftmargin=0mm\rightmargin=0cm}\small\item[]}{\endlist}

\begin{document}

\title{Trusty URIs: Verifiable, Immutable, and Permanent Digital Artifacts for Linked Data}

\author{
Tobias Kuhn\inst{1} and Michel Dumontier\inst{2}
}

\institute{
  Department of Humanities, Social and Political Sciences, ETH Zurich, Switzerland\\
\and
  Stanford Center for Biomedical Informatics Research, Stanford University, USA
\smallskip\\
  \texttt{tokuhn@ethz.ch},
  \texttt{michel.dumontier@stanford.edu}
}

\maketitle

\begin{abstract}
To make digital resources on the web verifiable, immutable, and permanent, we propose a technique to include cryptographic hash values in URIs. We call them \emph{trusty URIs} and we show how they can be used for approaches like nanopublications to make not only specific resources but their entire reference trees verifiable. Digital artifacts can be identified not only on the byte level but on more abstract levels such as RDF graphs, which means that resources keep their hash values even when presented in a different format. Our approach sticks to the core principles of the web, namely openness and decentralized architecture, is fully compatible with existing standards and protocols, and can therefore be used right away. Evaluation of our reference implementations shows that these desired properties are indeed accomplished by our approach, and that it remains practical even for very large files.
\end{abstract}

\section{Introduction}

The vision of the semantic web is to make the content of the web machine-interpretable, allowing, among other things, for automated aggregation and sophisticated search procedures over large amounts of linked data. As even human users are sometimes easy to trick by spam and fraudulent content that can be found on the web, we should be even more concerned in the case of automated algorithms that autonomously analyze semantic web content. Without appropriate counter-measures, malicious actors can sabotage or manipulate such algorithms by adding just a few carefully manipulated items to large sets of input data. To solve this problem, we propose an approach to make items on the (semantic) web verifiable, immutable, and permanent.
This approach includes cryptographic hash values in Uniform Resource Identifiers (URIs) and adheres to the core principles of the web, namely openness and decentralized architecture.

A cryptographic hash value (sometimes called \emph{cryptographic digest}) is a short random-looking sequence of bytes (or, equivalently, bits) that are calculated in a complicated yet perfectly predictable manner from a digital artifact such as a file. The same input always leads to exactly the same hash value, whereas just a minimally modified input leads to a completely different value. While there is an infinity of possible inputs that lead to a specific given hash value, it is impossible in practice (for strong state-of-the-art hash algorithms) to reconstruct \emph{any} of the possible inputs just from the hash value. This means that if you are given some input and a matching hash value, you can be sure that the hash value was obtained from exactly that input. On this basis, our proposed approach boils down to the idea that references can be made completely unambiguous and verifiable if they contain a hash value of the referenced digital artifact. Our method does not apply to all URIs, of course, but only to those that are meant to represent a specific and immutable digital artifact.

Let us have a look at a concrete example: Nanopublications have been proposed as a new way of scientific publishing \cite{groth2010isu}. The underlying idea is that scientific results should be published not just as narrative articles but in terms of minimal pieces of computer-interpretable results in a formal semantic notation (i.e. RDF). Nanopublications can cite other nanopublications via their URIs, thereby creating complex citation networks. Published nanopublications are supposed to be immutable, but the current web has no mechanism to enforce this: It is well-known that even artifacts that are supposed to be immutable tend to change over time, while often keeping the same URI reference. For approaches like nanopublications, however, it is important to specify exactly what version of what resource they are based on, and nobody should be given the opportunity to silently modify his or her already published contributions.

\begin{figure}[t]
\begin{center}
\includegraphics[width=\textwidth]{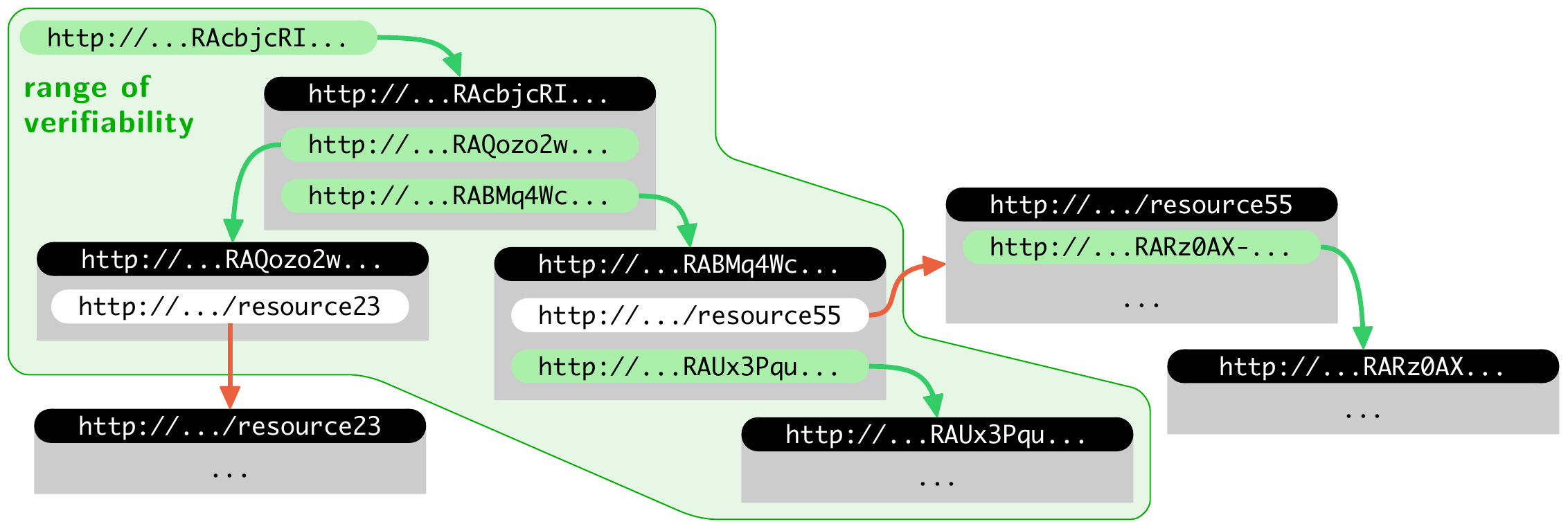}
\caption{Schematic illustration of the range of verifiability for the trusty URI on the top left. The green area shows its range of verifiability that covers all artifacts that can be reached by following trusty URI links (green arrows).}
\label{fig:verifiability}
\end{center}
\end{figure}

With the approach outlined below, nanopublications can be identified with \emph{trusty URIs} that include cryptographic hash values calculated on the RDF content. For example, let us assume that you have a nanopublication with identifier $I_1$ that cites another nanopublication with identifier $I_2$. If you want to find the content of $I_2$, you can simply search for it on the web, not worrying whether the source is trustworthy or not, and once you have found an artifact that claims to be $I_2$, you only have to check whether the hash value actually matches the content. If it does, you got the right nanopublication, and if not you have to keep searching (this can of course be automated). A trusty URI like $I_1$ does not only allow you to retrieve its nanopublication in a verifiable way, but in the next step also all nanopublications it cites (such as $I_2$) and all nanopublications they cite and so on. Any trusty URI in a way ``contains'' the complete backwards history.
In this sense, the ``range of verifiability'' of a resource with a trusty URI is not just the resource itself, but the complete reference tree obtained by recursively following all contained trusty URIs. This is illustrated in Figure \ref{fig:verifiability}.

Another important property of nanopublications is that they are self-con\-tained in the sense that they consist not only of the actual scientific assertions but also of their provenance information and meta-data. This means that nanopublications contain self-references in the form of their own identifying URIs. The calculation of a trusty URI must therefore allow for the resulting URI to be part of the digital artifact it is calculated on (this might sound impossible at first, but we show below how it can be achieved).
This leads us to the formulation of the following requirements for our approach:
\begin{enumerate}
\item To allow for verification of not only the given digital artifact but its entire reference tree, the hash should be part of the URI of the artifact.
\item To allow for the inclusion of meta-data, digital artifacts should be allowed to contain self-references (i.e. their own URIs).
\item The verification should be performed on a more abstract level than just the bytes of a file, with modules for different types of content. It should be possible to verify a digital artifact even if it is presented in a different format.
\item The complete approach should be decentralized and open: Everybody should be allowed to make verifiable URIs without a central authority.
\item The approach should be based on current established standards and be compatible with current tools and formats, so that it can be used right away.
\end{enumerate}
Though there are a number of related approaches, we are not aware of any general approach that complies with all these requirements. In particular, requirements 2 and 3 are not addressed by existing approaches.
The main benefits of artifacts with a trusty URI are that they are (1) verifiable, (2) immutable, and (3) permanent. Let us briefly explain what we mean by these properties.

Trusty URI artifacts are \emph{verifiable} in the sense that a retrieved artifact for a given URI can be checked to contain the content the URI is supposed to represent. It can be detected if the artifact got corrupted or manipulated on the way, assuming that the trusty URI for the required artifact is known, e.g. because another artifact contains it as a link. (Of course, somebody can give you a manipulated artifact with a \emph{different} trusty URI.)

It directly follows that trusty URI artifacts are \emph{immutable}, as any change in the content also changes its URI, thereby making it a \emph{new} artifact. Again, you can of course change your artifact \emph{and} its URI and claim that it has always been like this. You can get away with that if the trusty URI has not yet been picked up by third parties, i.e. linked by other resources. Once this is the case, you cannot change it anymore, because all these links will still point to the old trusty URI and everybody will notice that your new artifact is a different one.

Third, trusty URI artifacts are \emph{permanent} if we assume that there are search engines and web archives crawling the artifacts on the web and caching them. In this situation, any artifact that is available on the web for a sufficiently long time will remain available forever. If an artifact is no longer available in its original location (e.g. the one its URI resolves to), one can still retrieve it from the cache of search engines or web archives. The trusty URI guarantees that it is the artifact you are looking for, even if the location of the cached artifact is not trustworthy or it was cached from an untrustworthy source.

\section{Background}

There are a number of related approaches based on cryptographic hash values.
The Git version control system ({\small\url{git-scm.com}}), for example, uses hash values to identify commits of distributed repositories.
An important difference to our approach is that hash values (called \emph{checksums} in Git) are used to identify the respective artifacts (\emph{commits} in Git) only within a given repository and not on the web scale. A second important difference is that the hash represents the byte content of files, whereas our approach allows for digital content at different levels of abstraction. On the technical side, Git uses the SHA-1 algorithm, which is no longer considered secure (which is not a serious problem for Git, because typically only trusted parties have write access to a repository).

The proposed standard for Named Information (ni) URIs \cite{keranen2013ietf} is another important related approach. It introduces a new URI protocol \texttt{ni} to refer to digital artifacts with hash values in a uniform way. These are two examples of ni-URIs:
\begin{uriquote}
\url{ni:///sha-256;UyaQV-Ev4rdLoHyJJWCi11OHfrYv9E1aGQAlMO2X_-Q}\\
\url{ni://example.org/sha-256;5AbXdpz5DcaYXCh9l3eI9ruBosiL5XDU3rxBbBaUO70}
\end{uriquote}
The ni-URI approach allows for different hash algorithms, such as SHA-256 (which is, in contrast to SHA-1, considered secure) and optional specification of an authority, such as \texttt{example.org}, where the artifact can be found. It misses, however, some of the features of our requirements list. As with Git, ni-URIs do not define how digital artifacts can be represented at a more abstract level than their sequence of bytes, and self-references are not supported. Furthermore, current browsers do not recognize the \texttt{ni} protocol, and administrator access to a server is needed to make these URIs resolvable. The latter two points are not a real problem in the long run, but they might hinder the adoption of the standard in the first place. The approach presented in this paper is complementary and compatible. We propose trusty URIs, which can be mapped to ni-URIs but are more flexible and provide additional features.

There are a number of existing approaches to include hash values in URIs for verifiability purposes, e.g. for legal documents \cite{hoekstra2011iswc}. The downside of such custom-made solutions is that custom-made software is required to generate, resolve, and check the hash references.
Standards have been proposed for the verification of quantitative datasets \cite{altman2007dlib} and XML documents \cite{bartel2008xmldsig}, but they are not general enough to cover RDF content (at least not in a convenient way) and keep the hash value separate from the URI reference, which means that the range of verifiability does not directly extend to referenced artifacts.

To calculate hash values on content that is more abstract than just a fixed sequence of bytes, common approaches require the normalization (also called \emph{canonicalization}) of the respective data structures such as RDF graphs. In the general case, RDF graph normalization is known to be a very hard problem, possibly unsolvable in polynomial time \cite{carroll2003iswc}. This stems from the difficulty of handling blank nodes, i.e. identifiers that are only unique in a local scope and can be locally renamed without effects on semantics. Without blank nodes, normalization boils down to sorting of RDF triples, which can be performed in $O(n \log n)$.
The need for sorting can even be eliminated by using incremental cryptography \cite{bellare1994crypto}, which allows for calculating digests for RDF graphs without blank nodes in linear time \cite{sayers2004rdfdigets}. Such incremental approaches, however, are not as well-studied as mainstream cryptography methods, and open the possibility of new kinds of attacks \cite{phan2006cs}.
Efficient normalization algorithms that support blank nodes put restrictions on the graph structure and require additional (semantically neutral) triples to be added to some graphs before they can be processed \cite{carroll2003iswc,sayers2004rdfdigets}.

Similar methods to the ones presented in this paper, i.e. calculating hash values in a format-independent manner, have been proposed to track the provenance of data sets \cite{mccusker2012ieee}. This has been used to define a conceptualization of multi-level identities for digital works based on cryptographic digests and formal semantics, covering different conceptual levels from single HTTP transactions to high-level content identifiers \cite{mccusker2012ipaw}.

\section{Approach}

We propose here a modular approach, where different modules handle different kinds of content on different conceptual levels of abstraction, from byte level to high-level formalisms. Besides that, the most important features of our approach are self-references, the handling of blank nodes, and the mapping to ni-URIs.

\subsubsection{General Structure.}

Trusty URIs end with a hash value in Base64 notation (i.e. \texttt{A}--\texttt{Z}, \texttt{a}--\texttt{z}, \texttt{0}--\texttt{9}, \texttt{-}, and \texttt{\_} representing the numbers from 0 to 63) that is preceded by a module identifier.
This is an example:
\begin{uriquote}
\url{http://example.org/r1.RA5AbXdpz5DcaYXCh9l3eI9ruBosiL5XDU3rxBbBaUO70}
\end{uriquote}
Everything that comes after \texttt{r1.} is the part that is specific to trusty URIs, which we call \emph{artifact code}. Its first two characters \texttt{RA} identify the module specifying its type (first character) and version (second character). The remaining 43 characters represent the actual hash value. The modules defined so far use SHA-256 hashes, but future modules might use other hash functions. For convenience reasons, a file extension like \texttt{.nq} can be added to the end of such URIs:
\begin{uriquote}
\url{http://example.org/r1.RA5AbXdpz5DcaYXCh9l3eI9ruBosiL5XDU3rxBbBaUO70.nq}
\end{uriquote}
This is technically not a trusty URI anymore, but it is easy to strip the extension and get the respective trusty URI from it. As the hash is located in the final part of the URI, it is straightforward to store it in file names and to deal with it in a local file system without worrying about the first part of the URI:
\begin{quote}\small
\verb|r1.RA5AbXdpz5DcaYXCh9l3eI9ruBosiL5XDU3rxBbBaUO70.nq|
\end{quote}
We call these \emph{trusty files}.
The precise specification of trusty URIs can be found online.\footnote{\url{https://github.com/trustyuri/trustyuri-spec}}
As a general side remark, it is noteworthy that our approach entails a certain shift of authority: Once a trusty URI is established, its artifact code defines what object it refers to, and the issuing authority has no longer the power to change its meaning.

\subsubsection{Self-References.}

\newcommand{\hashplaceholder}{{\setlength{\fboxsep}{0.5mm}\hspace{0.2mm}\fbox{\normalfont\textsc{c}}\hspace{0.2mm}}}

To support self-references, i.e. resources that contain their own trusty URI, the generation process involves not just to compute the hash from a given artifact but to actually transform the artifact into a new version that contains the newly generated trusty URI.
For example, a resource like \texttt{\small{}http://example.org/r2} might have the following RDF content with a self-reference:
\begin{quote}\small
\texttt{<http://example.org/r2> dc:description "something" .}
\end{quote}
To transform such a resource, we first define the structure of the new trusty URI by adding a placeholder {\hashplaceholder} where the artifact code should eventually appear. In the given example, the content would then look like this:
\begin{quote}\small
\texttt{<http://example.org/r2.{\hashplaceholder}> dc:description "something" .}
\end{quote}
Note that it is necessary to add a non-Base64 character (in this case a dot ``\texttt{.}'') as a delimiter in front of {\hashplaceholder} if it would otherwise be preceded by a Base64 character. On such content, we can calculate a hash value by interpreting the placeholder {\hashplaceholder} as a blank space (the result is unambiguous as URIs are not allowed to otherwise contain blank spaces). Then we can replace the placeholder by the calculated artifact code and we end up with a trusty URI like this:
\begin{uriquote}
\url{http://example.org/r2.RAi7LA7Zlew99hdp0joN0APT4_uB3XDFwduiKXnNBja5E}
\end{uriquote}
For strong hashing algorithms, it is impossible in practice that this calculated sequence of bytes was already part of the original content before the transformation. This entails that the replacing of the placeholder is reversible.

This reversibility is needed once an existing trusty URI resource containing self-references should be verified. We can revert the transformation described above by replacing all occurrences of the artifact code with a blank space, and then calculate the hash in the same way as when a resource is transformed. The content is successfully verified if and only if the resulting hash matches the one from the trusty URI.

\subsubsection{Blank Nodes.}

The support for self-references requires us to transform the preliminary content of a trusty URI artifact into its final version, and we can make use of this transformation to also solve the problem of blank nodes in RDF.
Our approach is to eliminate blank nodes during the transformation process by converting them into URIs. Blank nodes can be seen as existentially quantified variables, which we can turn into constants by Skolemization, i.e. by introducing URIs that have not been used anywhere before.
Using the trusty URI with a suffix enumerating the blank nodes, we can create such URIs guaranteed to have never been used before (the artifact code being just a placeholder at first, as above):
\begin{uriquote}
\url{http://example.org/r3.RACjKTA5dl23ed7JIpgPmS0E0dcU-XmWIBnGn6Iyk8B-U..1}\\
\url{http://example.org/r3.RACjKTA5dl23ed7JIpgPmS0E0dcU-XmWIBnGn6Iyk8B-U..2}
\end{uriquote}
The two dots ``\texttt{..}'' indicate that these were derived from blank nodes, but they are now immutable URIs. This approach solves the problem of blank nodes for normalization, is completely general (i.e. works on any possible input graph), fully respects RDF semantics, and does not require auxiliary triples to be added.

\subsubsection{ni-URIs.}

Our approach is compatible with ni-URIs (see above), and all trusty URIs can be transformed into ni-URIs, with or without explicitly specifying an authority:
\begin{uriquote}
\url{ni:///sha-256;5AbXdpz5DcaYXCh9l3eI9ruBosiL5XDU3rxBbBaUO70}\\
\url{ni://example.org/sha-256;5AbXdpz5DcaYXCh9l3eI9ruBosiL5XDU3rxBbBaUO70}
\end{uriquote}
The fact that the module identifier is lost does not affect the uniqueness of the hash, but to verify a resource all available modules have to be tried in the worst case. To avoid this, we propose to use an optional argument \texttt{module}:
\begin{uriquote}
\url{ni:///sha-256;5AbXdpz5DcaYXCh9l3eI9ruBosiL5XDU3rxBbBaUO70?module=RA}
\end{uriquote}

\subsubsection{Modules.}

There are currently two module types available: \texttt{F} for representing byte-level file content and \texttt{R} for RDF graphs. For both types, version \texttt{A} is the only version available as of now, leading to the module identifiers \texttt{FA} and \texttt{RA}.
For module \texttt{FA}, a hash value is calculated using SHA-256 on the content of a file in byte representation.
The hash value is transformed to Base64 notation (after appending two zero-bits), and the resulting 43 characters make up the hash part of the trusty URI.
Module \texttt{RA} works on RDF content and can cover multiple named graphs. It supports self-references and handles blank nodes as described above. To calculate the hash, the RDF statements are sorted, then they are serialized in a given way (interpreting the artifact's hash as a blank space), and finally SHA-256 is applied in the same way as for \texttt{FA}.

Note that for an RDF document, either of the modules \texttt{FA} and \texttt{RA} could be used. The right choice depends on what the URI should identify. If it should identify a \emph{file} in a particular format and containing a fixed number of bytes, then \texttt{FA} should be used. If it should, however, identify \emph{RDF content} independently of its serialization in a particular file, then \texttt{RA} should be used.
For modules such as \texttt{RA} that operate not just on the byte level, \emph{content negotiation} can be used to return the same content in different formats (depending on the requested content type in the HTTP request) when a trusty URI is accessed.

Even though we focus on RDF in this paper, the approach and architecture of trusty URIs are general and we plan to provide modules for additional kinds of content in the future. This could include tabular or matrix content (e.g. CSV or Excel files), content with tree structure (e.g. XML), hypertext (e.g. HTML or Markdown), bitmaps (e.g. PNG or JPEG), and vector graphics (e.g. SVG). New modules might also become necessary if the used hash algorithms should become vulnerable to attacks in the future.

\section{Implementation}

There are currently three trusty URI implementations in the form of code libraries in Java, Perl, and Python.\footnote{\url{https://github.com/trustyuri/trustyuri-java},\hspace{1ex}\url{https://github.com/trustyuri/trustyuri-perl}, \url{https://github.com/trustyuri/trustyuri-python}}
The Java implementation uses the \emph{Sesame} library \cite{broekstra2002iswc} for RDF processing and the \emph{nanopub-java} library\footnote{\url{https://github.com/Nanopublication/nanopub-java}} for dealing with nanopublications. The Perl implementation makes use of the \emph{Trine} package for processing RDF, and the Python implementation uses the \emph{RDFLib} package.\footnote{\url{http://search.cpan.org/dist/RDF-Trine/}, \url{https://github.com/RDFLib/rdflib}}

These implementations provide a number of common functions for the different modules and formats. Currently, the following functions are available:
\begin{description}
\item[CheckFile] takes a file and validates its hash by applying the respective module.
\item[ProcessFile] takes a file, calculates its hash using module \texttt{FA}, and renames it to make it a trusty file.
\item[TransformRdf] takes an RDF file and a base URI, and transforms the file into a trusty file using module \texttt{RA}.
\item[TransformLargeRdf] is the same as above but using temporary files instead of loading the entire content into memory.
\item[TransformNanopub] takes a nanopublication file and calls TransformRdf to transform it.
\item[CheckLargeRdf] checks an RDF file using module \texttt{RA} without loading the whole content into memory but using temporary files instead.
\item[CheckSortedRdf] checks an RDF file assuming that it is already sorted (and raises an error otherwise). The current implementations generate such sorted files by default, but this is not required by the specification.
\item[CheckNanopubViaSparql] takes a SPARQL endpoint URL and a trusty URI representing a nanopublication, retrieves the nanopublication from the repository, and tries to validate it.
\item[RunBatch] reads commands (any of the above) from a file and executes them one after the other.
\end{description}
Not all these functions are currently supported by all implementations, as shown in Table \ref{tab:trustyurilibs}.

\begin{table}[t]
\begin{center}\small
\begin{tabular}{l|ll|ccc}
module & function & format & Java & Perl & Python \\
\hline
\emph{(general)} & RunBatch & & \checkmark & \checkmark & \checkmark \\
\hline
\multirow{2}{*}{File} & CheckFile & & \checkmark & \checkmark & \checkmark \\
\cline{2-6}
 & ProcessFile & & \checkmark & \checkmark & \checkmark \\
\hline
\multirow{16}{*}{RDF} & \multirow{6}{*}{CheckFile} & RDF/XML & \checkmark & \checkmark & \checkmark \\
 & & Turtle & \checkmark & \checkmark & \checkmark \\
 & & N-Triples & \checkmark & \checkmark & \checkmark \\
 & & TriX & \checkmark & -- & \checkmark \\
 & & TriG & \checkmark & \checkmark & -- \\
 & & N-Quads & \checkmark & \checkmark & \checkmark \\
\cline{2-6}
 & CheckLargeRdf & \emph{(all of the above)} & \checkmark & & \\
\cline{2-6}
 & CheckSortedRdf & \emph{(all of the above)} & \checkmark & & \\
\cline{2-6}
 & \multirow{6}{*}{TransformRdf} & RDF/XML & \checkmark & & \checkmark \\
 & & Turtle & \checkmark & & \checkmark \\
 & & N-Triples & \checkmark & & \checkmark \\
 & & TriX & \checkmark & -- & \checkmark \\
 & & TriG & \checkmark & -- & -- \\
 & & N-Quads & \checkmark & & \checkmark \\
\cline{2-6}
 & TransformLargeRdf & \emph{(all of the above)} & \checkmark & & \\
\cline{2-6}
 & \multirow{3}{*}{TransformNanopub} & TriX & \checkmark & -- & \\
 & & TriG & \checkmark & -- & -- \\
 & & N-Quads & \checkmark & & \\
\cline{2-6}
 & CheckNanopubViaSparql & & \checkmark & & \\
\hline
\end{tabular}
\end{center}
\caption{Comparison of the different trusty URI libraries (`\checkmark' = implemented features; `--' = cases where the necessary features are not available in the used RDF libraries)}
\label{tab:trustyurilibs}
\end{table}

\section{Application}

Below, we describe two applications of the trusty URI approach: one involving nanopublications (nanobrowser) and one involving a dataset in RDF format with a large variation in file size (Bio2RDF).

\subsection{Nanobrowser}

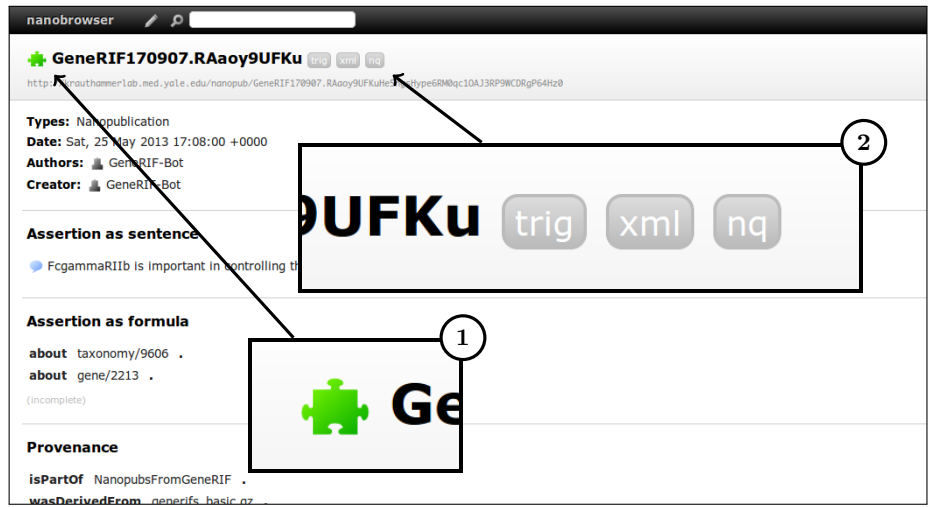
\begin{figure}[t]
\begin{center}
\setlength{\fboxsep}{0mm}
\begin{tikzpicture}
\node[inner sep=0mm] (img) at (0,0) {\fbox{\includegraphics[width=0.999\textwidth]{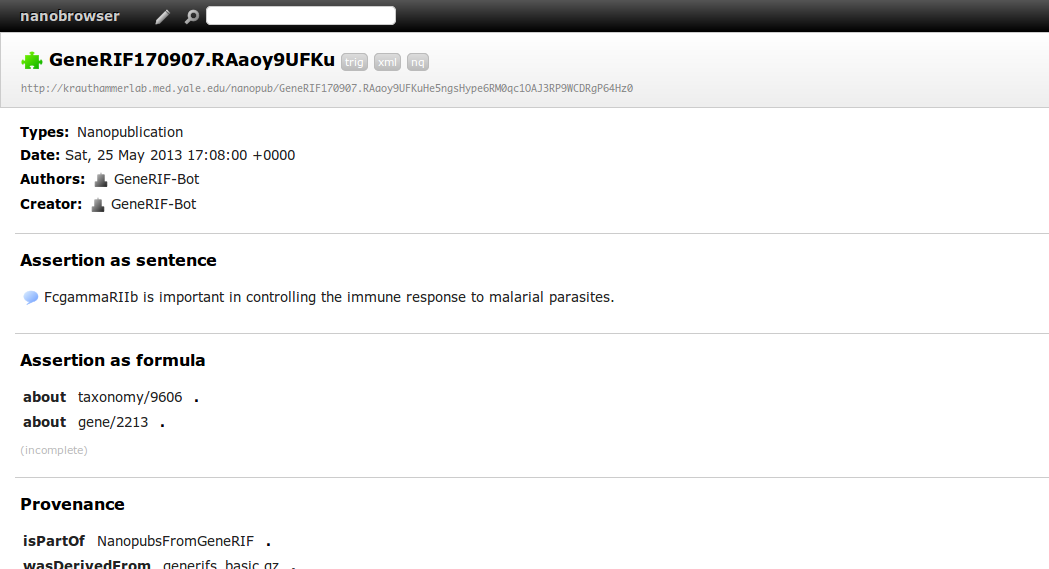}}};
\node[inner sep=0mm] (i) at (-1.5,-2) {\setlength{\fboxrule}{0.5mm}\fbox{\includegraphics[scale=0.4]{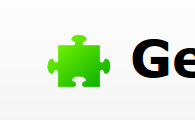}}};
\node[circle,fill=white,draw,very thick] at (i.north east) {\textbf{1}};
\draw[->,very thick] (i) -- (-5.5,2.4);
\node[inner sep=0mm] (l) at (1.5,0.5) {\setlength{\fboxrule}{0.5mm}\fbox{\includegraphics[scale=0.4]{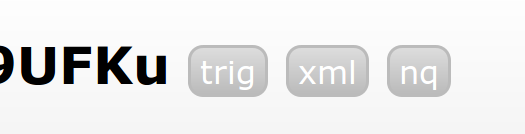}}};
\node[circle,fill=white,draw,very thick] at (l.north east) {\textbf{2}};
\draw[->,very thick] (l) -- (-1,2.4);
\end{tikzpicture}
\caption{Screenshot of the nanobrowser interface. A green jigsaw puzzle icon (1) indicates successful verification of nanopublications, which can be downloaded (and verified locally) in different formats (2).}
\label{fig:nanobrowser}
\end{center}
\end{figure}

Nanobrowser \cite{kuhn2013eswc} is a prototype of a web application via which nanopublications can be searched, browsed, published, and commented. Figure \ref{fig:nanobrowser} shows a screenshot. Nanobrowser applies a number of extensions to the nanopublication approach, such as support for semi-formal and informal statements (represented by atomic and independent English sentences, i.e. a kind of controlled natural language \cite{kuhn2014cl}) and support for meta-nanopublications, e.g. nanopublications containing opinions on other nanopublications.

All nanopublications created via the nanobrowser interface are identified by trusty URIs. If a user requests a nanopublication with a given trusty URI, it is retrieved from the internal triple store and verified before it is shown to the user. A green jigsaw puzzle icon indicates that the verification was successful (see Figure \ref{fig:nanobrowser}). A particular nanopublication can be downloaded in different formats and its trusty URI can be checked locally and independently of the format.

\subsection{Bio2RDF}

Bio2RDF ({\small\url{bio2rdf.org}}) is an open-source project focused on the provision of linked data for the life sciences \cite{callahan2013eswc,belleau2008biomedinform}. Bio2RDF scripts convert heterogeneously formatted data (e.g. flat files, tab-delimited files, dataset-specific formats, SQL, and XML) into a common format --- RDF. Bio2RDF entities are identified using URIs that are resolvable using the Bio2RDF Web Application, a servlet that answers HTTP requests by formulating SPARQL queries against the appropriate SPARQL endpoints. Over 1 billion triples for 19 resources were made available in the second coordinated release of Bio2RDF \cite{callahan2013eswc}, and mappings to the Semanticscience Integrated Ontology \cite{callahan2013jbiomedsem} have been established. Together, these serve to provide ontology-based access to data on the emerging semantic web.

The release numbers of Bio2RDF provide a way to refer to a specific version of a dataset, e.g. for citing it in a scientific article or a nanopublication. However this assumes trust in the Bio2RDF developers that they do not silently change the data of a particular release. Furthermore, an intruder might be able to change parts of the data without being noticed, the data might get corrupted or manipulated when transferred or downloaded, and there might be no other trusted parties providing the dataset if the Bio2RDF website should become temporarily or permanently inaccessible. The use of trusty URIs would solve all these problems. Below we show an evaluation on release 2 of Bio2RDF, and we plan to provide trusty URIs for the datasets of its upcoming next release.

\section{Evaluation}

Below we present some experiments on the trusty URI concept and its implementations, based on two collections of RDF files.

\begin{table}[tp]
\begin{center}
{\scriptsize\sffamily
\renewcommand{\arraystretch}{1.75}
\textsc{Normal Mode}\smallskip\\
\begin{tabular}{r|lr|rrrrc|rrr}
\hline
 & \multicolumn{2}{c|}{method} & \multicolumn{5}{c|}{time in seconds} & \multicolumn{3}{c}{result} \\
 & impl. & format & mean & stdev & min & max & histogram & valid & invalid & error \\
\hline
\multirow{7}{*}{\begin{sideways}\phantom{p}valid files\phantom{p}\end{sideways}}
 & Java & N-Quads & 0.5229 & 0.0591 & 0.3750 & 5.5420& \multirow{7}{*}{\includegraphics[trim=5mm 13mm 1mm 11mm, clip=true, scale=0.49]{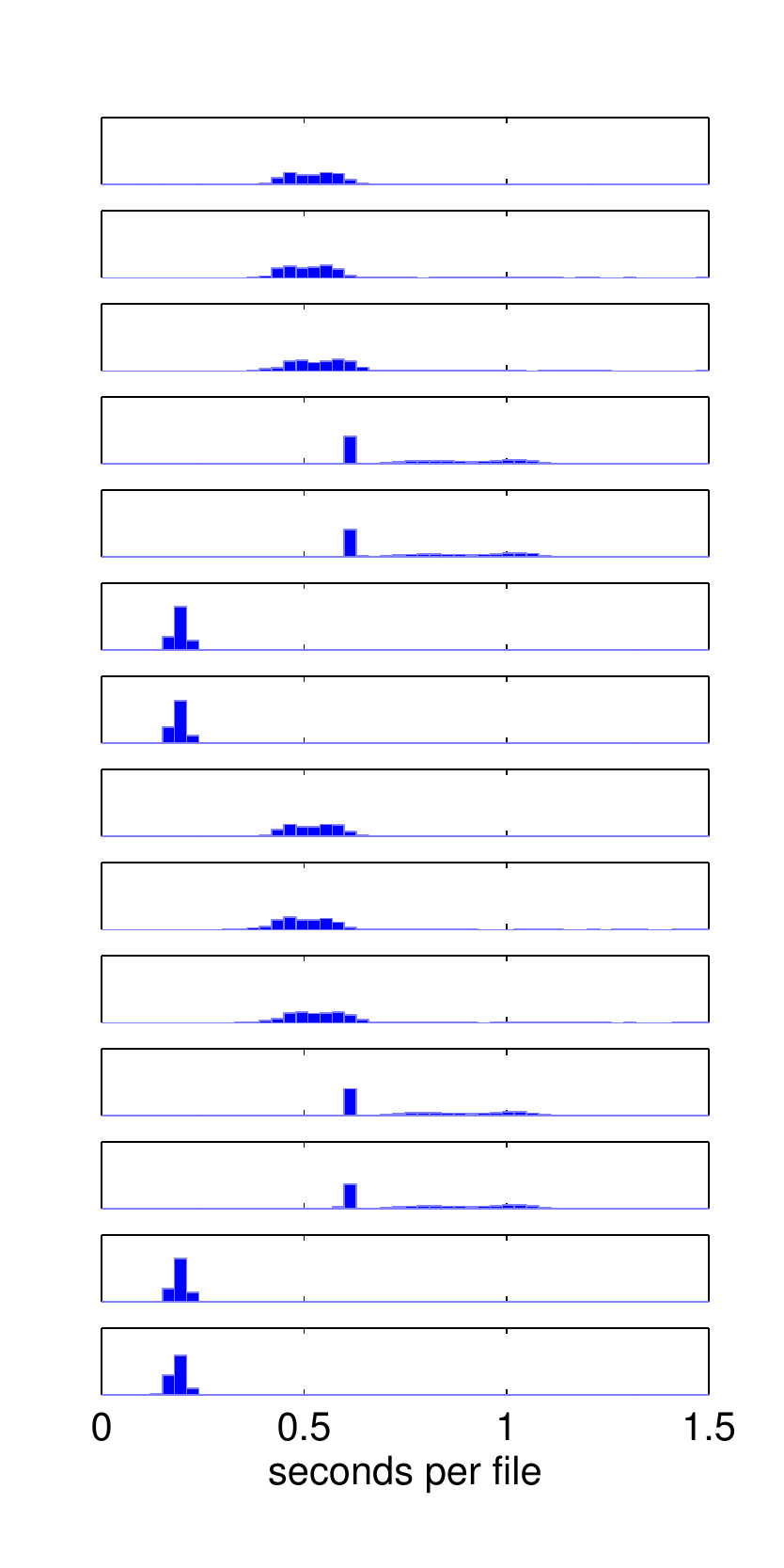}} & 100\% & 0\% & 0\% \\
 & Java & TriG & 0.5113 & 0.0569 & 0.3650 & 5.5340 & & 100\% & 0\% & 0\% \\
 & Java & TriX & 0.5383 & 0.0648 & 0.3900 & 5.5240 & & 100\% & 0\% & 0\% \\
 & Perl & N-Quads & 0.7843 & 0.1713 & 0.5990 & 5.7960 & & 100\% & 0\% & 0\% \\
 & Perl & TriG & 0.7901 & 0.1734 & 0.6030 & 5.7840 & & 100\% & 0\% & 0\% \\
 & Python & N-Quads & 0.1935 & 0.0164 & 0.1150 & 0.3050 & & 100\% & 0\% & 0\% \\
 & Python & TriX & 0.1912 & 0.0162 & 0.1190 & 0.3460 & & 100\% & 0\% & 0\% \\
\hline
\multirow{7}{*}{\begin{sideways}corrupted files\end{sideways}}
 & Java & N-Quads & 0.5227 & 0.0591 & 0.3450 & 5.5420 & & 0\% & 99.72\% & 0.28\% \\
 & Java & TriG & 0.5003 & 0.0621 & 0.3200 & 5.4250 & & 0\% & 83.37\% & 16.63\% \\
 & Java & TriX & 0.5322 & 0.0655 & 0.3360 & 5.5230 & & 0.83\% & 84.15\% & 15.03\% \\
 & Perl & N-Quads & 0.7842 & 0.1712 & 0.6000 & 5.8880 & & 0\% & 100\% & 0\% \\
 & Perl & TriG & 0.7872 & 0.1727 & 0.5700 & 5.8230 & & 0\% & 84.49\% & 15.51\% \\
 & Python & N-Quads & 0.1934 & 0.0165 & 0.1200 & 0.3080 & & 0\% & 100\% & 0\% \\
 & Python & TriX & 0.1884 & 0.0176 & 0.1070 & 0.2760 & & 0.12\% & 84.46\% & 15.42\% \\
\end{tabular}
\vspace{6mm}\\
\textsc{Batch Mode}\smallskip\\
\begin{tabular}{r|lr|rrrrc|rrr}
\hline
 & \multicolumn{2}{c|}{method} & \multicolumn{5}{c|}{time in seconds} & \multicolumn{3}{c}{result} \\
 & impl. & format & mean & stdev & min & max & histogram & valid & invalid & error \\
\hline
\multirow{7}{*}{\begin{sideways}\phantom{p}valid files\phantom{p}\end{sideways}}
 & Java & N-Quads & 0.0019 & 0.0062 & 0.0013 & 1.7202 & \multirow{7}{*}{\includegraphics[trim=5mm 13mm 1mm 11mm, clip=true, scale=0.49]{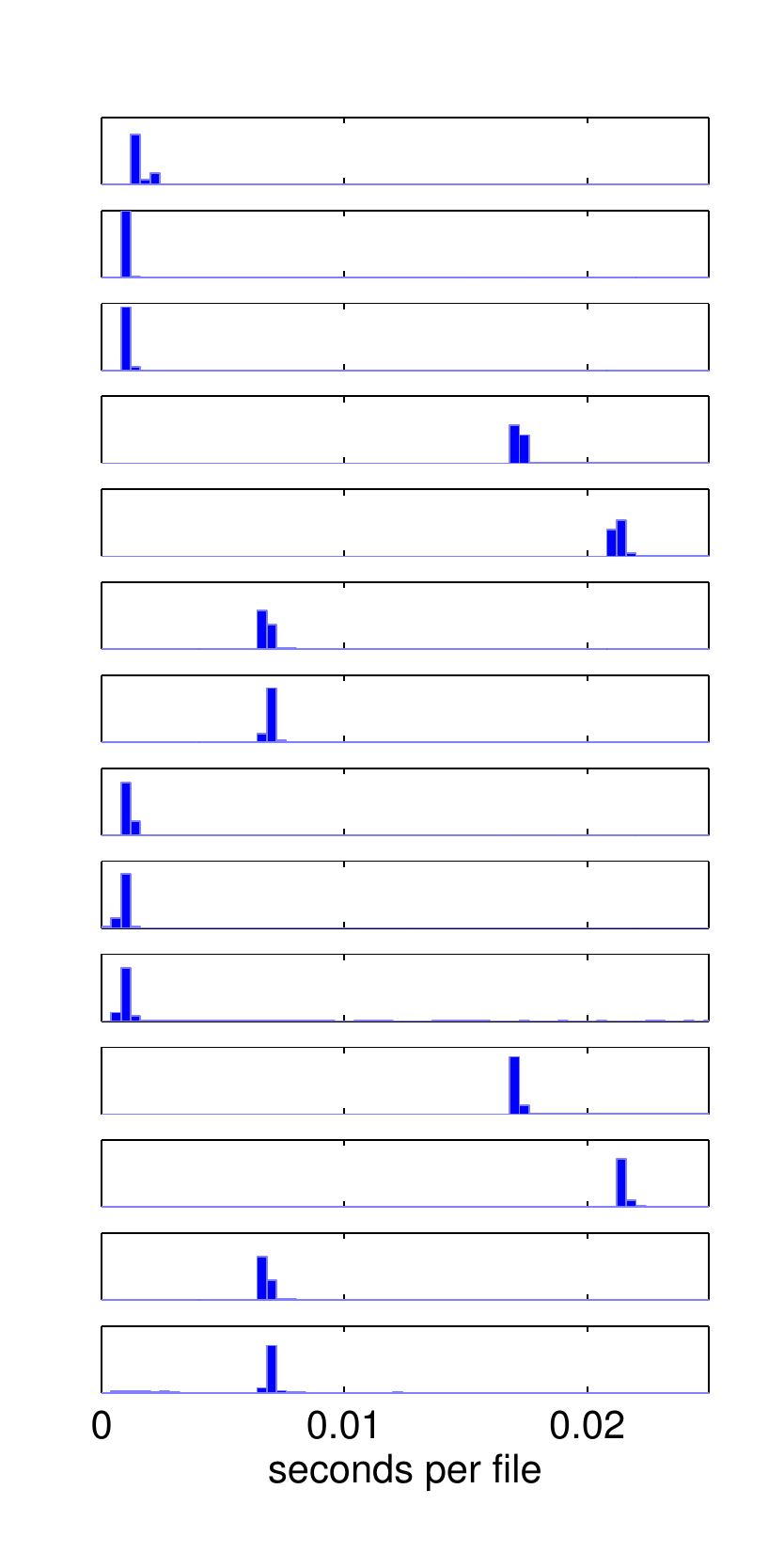}} & 100\% & 0\% & 0\% \\
 & Java & TriG & 0.0009 & 0.0050 & 0.0008 & 1.7412 & & 100\% & 0\% & 0\% \\
 & Java & TriX & 0.0011 & 0.0050 & 0.0009 & 1.5656 & & 100\% & 0\% & 0\% \\
 & Perl & N-Quads & 0.0172 & 0.0006 & 0.0171 & 0.0679 & & 100\% & 0\% & 0\% \\
 & Perl & TriG & 0.0214 & 0.0016 & 0.0211 & 0.0872 & & 100\% & 0\% & 0\% \\
 & Python & N-Quads & 0.0070 & 0.0011 & 0.0065 & 0.0644 & & 100\% & 0\% & 0\% \\
 & Python & TriX & 0.0070 & 0.0009 & 0.0066 & 0.0578 & & 100\% & 0\% & 0\% \\
\hline
\multirow{7}{*}{\begin{sideways}corrupted files\end{sideways}}
 & Java & N-Quads & 0.0012 & 0.0062 & 0.0006 & 1.6559 & & 0\% & 99.72\% & 0.28\% \\
 & Java & TriG & 0.0010 & 0.0049 & 0.0003 & 1.6335 & & 0\% & 83.37\% & 16.63\% \\
 & Java & TriX & 0.0011 & 0.0044 & 0.0005 & 1.3451 & & 0.83\% & 84.15\% & 15.03\% \\
 & Perl & N-Quads & 0.0171 & 0.0005 & 0.0169 & 0.0732 & & 0\% & 100\% & 0\% \\
 & Perl & TriG & 0.0195 & 0.0055 & 0.0007 & 0.0841 & & 0\% & 84.49\% & 15.51\% \\
 & Python & N-Quads & 0.0069 & 0.0011 & 0.0065 & 0.1716 & & 0\% & 100\% & 0\% \\
 & Python & TriX & 0.0063 & 0.0021 & 0.0006 & 0.1325 & & 0.12\% & 84.46\% & 15.42\% \\
\end{tabular}
\vspace{5mm}}
\caption{Performance and results of the different implementations for checking trusty URI nanopublications in normal mode (top) and batch mode (bottom) on valid and corrupted files.}
\label{tab:checkrdf}
\end{center}
\end{table}

\subsection{Hash Generation and Checking on Nanopublications}

To test our approach and to evaluate its implementations, we first took a collection of 156,026 nanopublications in TriG format that we had produced in previous work \cite{kuhn2013eswc}. We transformed these nanopublications into the formats N-Quads and TriX using existing off-the-shelf converters. Then, we transformed these into trusty URI nanopublications using the function TransformNanopub of the Java implementation. To be able to check not only positive cases (where checking succeeds) but also negative ones (where checking fails), we made copies of the resulting files where we changed a random single byte in each of them (only considering letters and numbers, and never replacing an upper-case letter by its lower-case version or vice versa, as some keywords are not case-sensitive). The resulting six sets of 156,026 files each (three formats, each in two versions: valid and corrupted) were the basis for our evaluation.

The first important result is that all original nanopublications ended up with the same trusty URI, no matter which format was used. This shows that our implementations are successful in handling the content on a more abstract level (i.e. RDF graphs in this case) leading to identical hash values for files that contain the same content but are quite different on the byte level.

Next, we checked the trusty URI of each nanopublication file with the function CheckFile of all implementations that support the respective format. The three right-most columns of Table \ref{tab:checkrdf} show the results. For all valid files (i.e. those we did not corrupt), all implementations correctly verified their trusty URIs. For the corrupted ones, where we randomly changed one byte, the checks almost always failed (by either calculating a different hash value than the one of the trusty URI, or by raising an error that the respective file was not well-formed).

The only corrupted files that were successfully validated were 1,290 TriX files (0.83\%) when running the Java implementation and 181 TriX files (0.12\%) when running the Python implementation. Looking at these concrete cases reveals that they are all harmless. In these cases, the randomly changed byte was not part of the RDF content, but of the meta-information. Due to minor bugs in the used RDF libraries, this meta-information is not sufficiently checked, which leads to accepting the valid content instead of failing because of violated well-formedness.
All our TriX files start with the following two lines:
\begin{quote}\footnotesize
\texttt{<?xml version='1.0' encoding='UTF-8'?>\\
<TriX xmlns='http://www.w3.org/2004/03/trix/trix-1/'>}
\end{quote}
The RDF implementations in Java and Python (or the respective system utilities to load XML files) do not properly check these two lines containing meta-data. Both libraries raise no error if a file starts with something like \texttt{<?Aml} instead of \texttt{<?xml} (106 files); the Python library accepts invalid XML version numbers such as \texttt{1.a} (73 files); and the Java library does not check the TriX namespace argument, raising no error if the argument name is changed to something like \texttt{xmlnZ} (175 files) or the URI is wrong, such as \texttt{.../Prix-1/} (1007 files). In addition, both libraries correctly accept the rare cases (2 files) where the XML version was changed from \texttt{1.0} to \texttt{1.1}, which is the only other valid XML version as of now (though much less common).

\subsection{Performance Tests on Nanopublications}

Next, we used the same set of nanopublication files to test the performance of the different modules for checking trusty URI artifacts in different formats. There are two scenarios of how to run such checks: One can run one after the other, as when a small number of nanopublications are manually checked, or one can execute such checks in the form of a batch job in a single program run, which is the preferred procedure to run a large number of checks without supervision. The time required per file is typically much lower in batch mode, as the runtime environment has to start and finalize only once. Therefore it makes sense to have a look at both scenarios.

Table \ref{tab:checkrdf} shows the results of these performance checks for the normal mode (top) and batch mode (bottom).
These results and the ones presented below were obtained on a Linux server (Debian) with 16 Intel Xeon CPUs of 2.27GHz and 24GB of memory.
As expected, the times are much lower in batch mode, but checking is reasonably fast also in normal mode. All average values are below 0.8s (0.03s for batch mode). Using Java in batch mode even requires only 1ms per file. Apart from the runtimes, the two modes had no effect on the results.

\subsection{Performance Tests on Bio2RDF}

The tests above cover only very small RDF files, but our approach should also work for larger files. For that reason, we performed a second evaluation on Bio2RDF, which includes much larger files. Release 2 of this dataset contains 874 RDF files in N-Triples format, but 16 of them lead to well-formedness errors when loaded with the current version of the Sesame library. (These problems might be related to the transition to the new RDF 1.1 standard, and they will be fixed for the next release of Bio2RDF.) This leaves us with 858 files of sizes ranging from 1.4kB to 177GB.

\begin{figure}[t]
\begin{center}
\includegraphics[trim=15mm 11.5mm 15mm 3mm, clip=true, width=\textwidth]{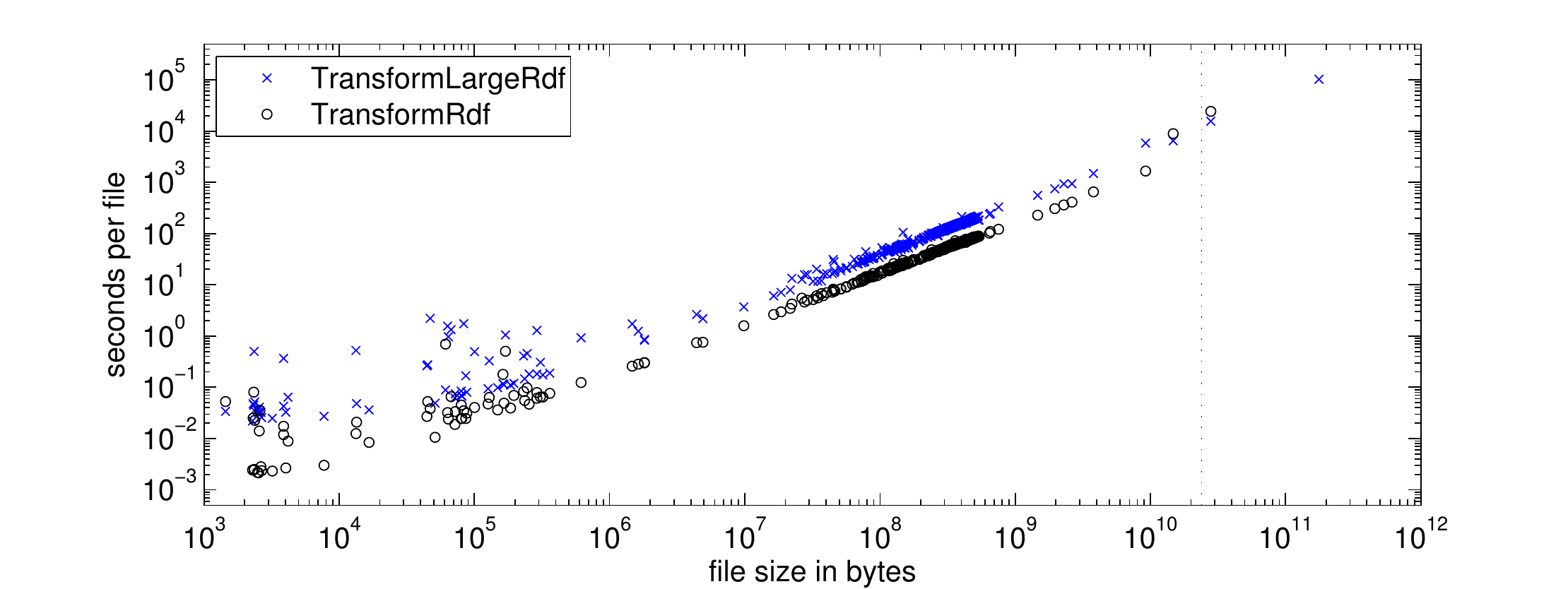}\\
\includegraphics[trim=15mm 0mm 15mm 6mm, clip=true, width=\textwidth]{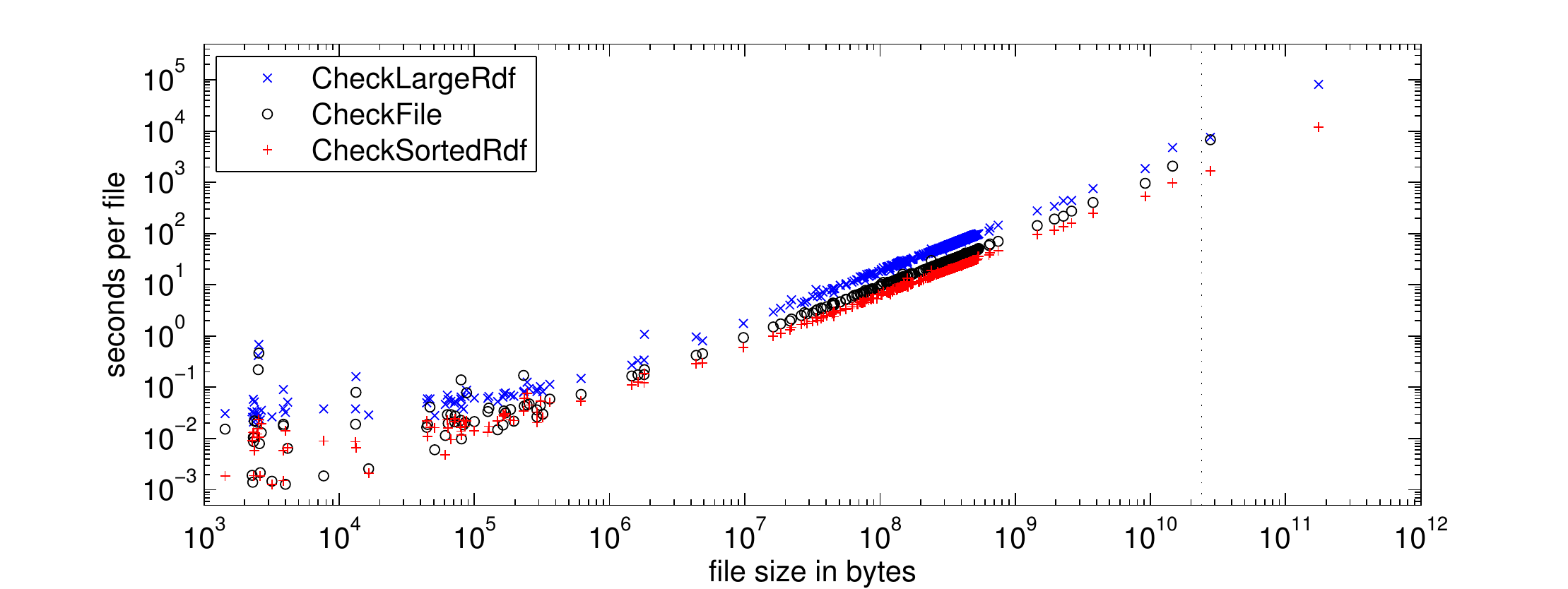}
\caption{Time required for transforming (top) and checking (bottom) files versus file size for the Bio2RDF dataset. The dotted line shows the available memory.
}
\label{fig:bio2rdf-time}
\end{center}
\end{figure}

Figure \ref{fig:bio2rdf-time} shows the results of these performance tests. There is a lot of random variation on the lower end, where files are smaller than 10MB and require less than three seconds to be processed. For the upper part, time values nicely follow near-linear trajectories (for the functions that do not load the whole content into memory). When hash calculation involves statement sorting, there is a strict theoretical limit on its performance due to the computational complexity of $O(n \log n)$.
TransformLargeRdf and CheckLargeRdf are superior to their counterparts only for very large files, and CheckSortedRdf is, as expected, faster than the other checking procedures. A large file of 2GB requires about five minutes to be transformed and about two minutes to be checked. Files larger than available memory take more time, but even the largest file of the dataset of 177GB was successfully transformed in 29 hours and checked in about three hours.

\section{Conclusions}

We have presented a proposal for unambiguous URI references to make digital artifacts on the (semantic) web verifiable, immutable, and permanent. If adopted, it could have a considerable impact on the structure and functioning of the web, could improve the efficiency and reliability of tools using web resources, and could become an important technical pillar for the semantic web, in particular for scientific data, where provenance and verifiability are crucial.
To improve reproducibility, for example, scientific data analyses might be conducted in the future within ``data projects'' analogous to today's software projects. The dependencies in the form of datasets could be automatically fetched from the web, similar to what Apache Maven ({\small\url{maven.apache.org}}) does for software projects but decentralized and verifiable.
In general, trusty URIs might contribute in a significant way to shape the future of publishing on the web.

\bibliography{trustyuris}
\bibliographystyle{plain}

\end{document}